# Fabrication of GaN/AlGaN 1D photonic crystals designed for nonlinear optical applications

T. Stomeo*[a], G. Epifani[a], V. Tasco[a], I. Tarantini[a], A. Campa[a], M. De Vittorio[a] and A. Passaseo[a], M. Braccini[b], M.C. Larciprete[b], C. Sibilia[b], F. A. Bovino[c]

[a] National Nanotechnology Laboratory of CNR-NANOSCIENCE, Distretto Tecnologico, Via Arnesano, 73100 Lecce, ITALY
[b] Dipartimento di Energetica-Università di Roma "La Sapienza", Via A. Scarpa 16, 00161 Roma, ITALY
[c] Quantum Optics Laboratory, Selex-SI, Via Puccini 2, 16154 Genova, ITALY

**ABSTRACT**

In this paper we present a reliable process to fabricate GaN/AlGaN one dimensional photonic crystal (1D-PhC) microcavities with nonlinear optical properties. We used a heterostructure with a GaN layer embedded between two Distributed Bragg Reflectors consisting of AlGaN/GaN multilayers, on sapphire substrate, designed to generate a $\lambda = 800$ nm frequency down-converted signal ($\chi^{(2)}$ effect) from an incident pump signal at $\lambda = 400$ nm. The heterostructure was epitaxially grown by metal organic chemical vapour deposition (MOCVD) and integrates a properly designed 1D-PhC grating, which amplifies the signal by exploiting the double effect of cavity resonance and non linear GaN enhancement. The integrated 1D-PhC microcavity was fabricate combing a high resolution e-beam writing with a deep etching technique. For the pattern transfer we used ~ 170 nm layer Cr metal etch mask obtained by means of high quality lift-off technique based on the use of bi-layer resist (PMMA/MMA). At the same time, plasma conditions have been optimized in order to achieve deeply etched structures (depth over 1 micron) with a good verticality of the sidewalls (very close to 90°). Gratings with well controlled sizes (periods of 150 nm, 230 nm and 400 nm respectively) were achieved after the pattern is transferred to the GaN/AlGaN heterostructure.

## 1. INTRODUCTION

The research towards compact and efficient nonlinear (NL) photonic devices has greatly improved in recent years in view of potential application in integrated photonics. Given the general structure of the NL interaction between an electrical field and the material response, in terms of polarization, one may act on susceptibilities $\chi$, local field and propagation in order to enhance a particular NL process. This translate into a research oriented towards new materials and new geometries. In particular, novel effects can be obtained when material properties are modulated at the sub-wavelength scale, that is by exploiting nanostructures.

The large bandgap group-III Nitride semiconductors (such as GaN, AlN and AlGaN) are promising nonlinear materials, due to their non-centro symmetric structure, showing a second order nonlinear optical responses $\chi^{(2)}$[1] comparable to conventional nonlinear crystals such as KTP or $LiNbO_3$ (in the visible range). Moreover these materials have a wide electronic band gap without absorption either of the fundamental wave in the near infrared or of the second-harmonic in the near UV, and high optical damage threshold. However, the efficiency of the nonlinear interaction in bulk GaN is too low for practical applications, because GaN is a highly dispersive material with low birefringence that prevents the realisation of phase matching in bulk samples.

*tiziana.stomeo@unisalento.it; phone +39-0832 298137; fax +39-0832 298386

This problem can be solved by combining the non linear properties of the nitride compounds with geometrically engineered photonic crystal (PhC) structures, tailoring exact phase-matching conditions and, with a periodic index variation, simultaneous field confinement. Thus, the novelty of our work is based on the development of a new class of PhC systems based on nitride heterostructures and their application for the study and the tailoring of NL optical effects.

One of most critical issues that must be faced in the fabrication of PhC devices in GaN/AlGaN high strained systems concerns the accurate control of geometrical parameters and sidewalls verticality. We developed a reliable technology to fabricate GaN/AlGaN 1D-PhC microcavities with high aspect ratio (15:1) for the propagation of coherent light generated by non linear optical processes. We used a sample with a GaN layer embedded between two (Distributed Bragg Reflectors) DBRs consisting of AlGaN/GaN multilayers, on sapphire substrate, designed to generate a $\lambda= 800$ nm frequency down-converted signal from an incident pump signal at $\lambda= 400$ nm. The heterostructure was epitaxially grown by metal organic chemical vapour deposition (MOCVD) technique and integrates a properly designed 1D-PhC grating, which amplifies the signal by exploiting the double effect of cavity resonance and non linear light enhancement. The 1D-PhC microcavity was fabricated combing a high resolution e-beam writing with a deep Inductively Coupled Plasma (ICP) etching in $SiCl_4$. The fabrication critical parameters were also analyzed by FDTD modelling tools in order to achieve an efficient propagation of coherent light generated by non linear optical processes.

## 2. DESIGN OF GAN/ALGAN 1D PHOTONIC CRYSTALS

The design of the vertical GaN/AlGaN 1D-PhC microcavity has been performed by exploiting a computer code based on finite difference time domain (FDTD) method. We used a heterostructure with a GaN layer embedded between two DBRs consisting of AlGaN/GaN multilayers, on sapphire substrate, designed to generate a $\lambda= 800$ nm frequency down-converted signal ($\chi^{(2)}$ effect) from an incident pump signal at $\lambda= 400$ nm, compatible with the optical set up used for the non linear measurements. GaN and AlN have not a high refractive index contrast ($n$AlN = 2.1, $n$GaN = 2.4 at a reference wavelength of 400 nm), therefore a high Al content layer has been chosen as low index material. The number of periods was fixed at 5 for both mirrors in order to have a fully epitaxial structure where the integration of a planar photonic crystal can be straightforward. Figure 1-a shows a s sketch of the basic structure.

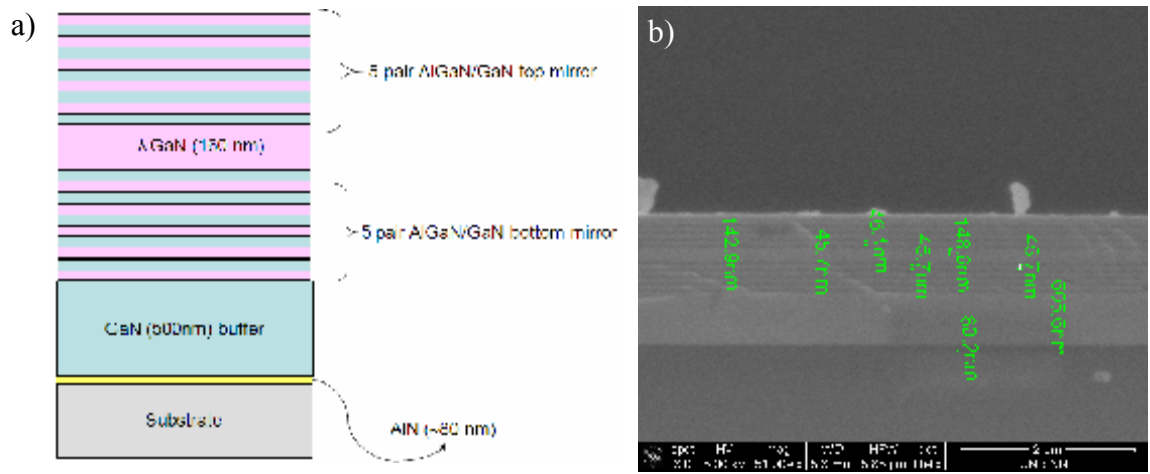

Figure 1. Sketch of the basic structure (a) and SEM cross-section of the grown sample (b).

Different 1D-PhC configurations have been analyzed and refined considering the technological issues and the optical (linear and non linear) measurements.

We started our analysis by considering a simplified geometry (fig. 1-a) consisting of a 2D model of a GaN slab waveguide imbedded in the microcavity structure followed by a grating with a central defect representing the output cavity. The purpose of this geometry was to enhance a pumping vertical field of 400 nm in order to get SPDC in the horizontal plane and to transmit the down converted photons in plane through a 1D grating to a defect designed for signal frequency conversion and extraction. The lateral size of the source block was designed to be of 0.5 and 1 μm, respectively, whereas the period of the grating was chosen to be of 400 nm. The defect slot was designed with a lateral size of 500 nm. Such a geometry was designed to provide an enhancement of the transmitted signal at the defect position as shown by the simulation in fig. 1-b.

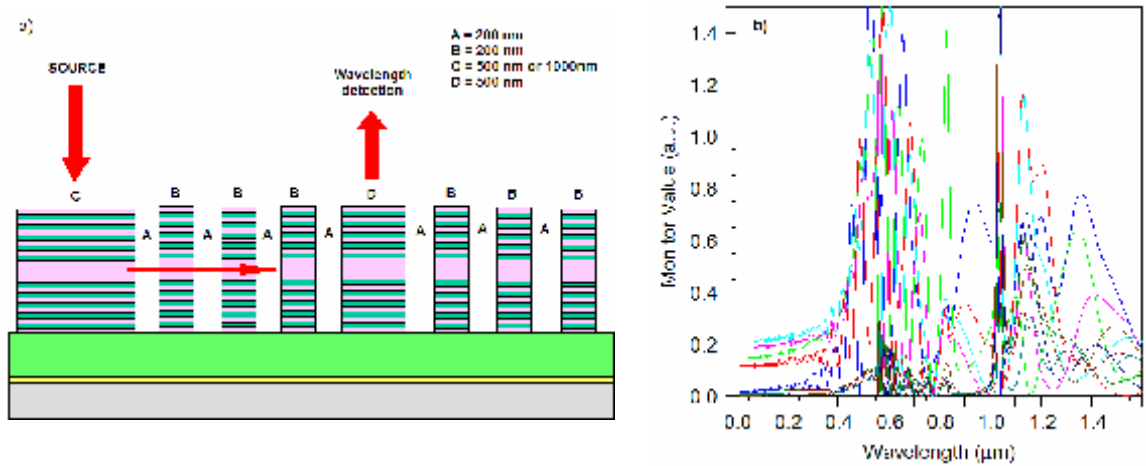

Figure 1. Simplified 1D PhC designed for the vertical GaN/AlGaN microcavity (a), and FDTD simulation of its optical response for the transmission of 800 nm signal (b).

The preliminary non linear measurements showed some problems in the vertical configurations, related to a bright and wide sapphire substrate fluorescence covering the down converted signal. As consequence, other geometries requiring lateral pumping experimental configurations have been considered and refined. These configurations should avoid the high noise induced by the substrate fluorescence. The structure that was then proposed is quite similar to the first one: also in this case a planar 1D-PhC is realised by a central block inserted in a grating with period of 230 nm (resonating at 400 nm pump wavelength), as shown in fig. 2. The slots constituting the grating are defined in the microcavity sample. Such a structure can be used in two different non linear experiments. In the first case (fig. 2), a pump signal in the y-z plane at 400 nm is resonant in the z direction and gets through the grating in the defect slot, where, in the vertical plane (x-z) it is coincident with the resonant mode. The $d_{33}$ high value of GaN would allow the generation of spontaneous parametric down conversion (SPDC) photons at 800 nm in the y-z plane. From these calculations it was also determined that, for future further improvements of the proposed optical device, to better optimize the phase matching conditions in this configuration the thickness of the grown GaN cavity should be designed for a resonance at 800 nm and then increased with respect to the considered microcavity sample.

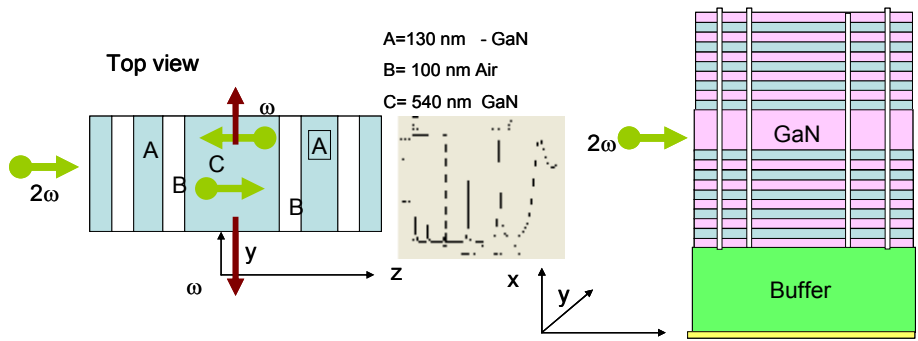

Figure. 2. Scheme of the optimised structure in the configuration of SPDC experiment.

The second proposed experiment was the use of this layout for studying second harmonic generation (SHG) processes. In particular, two pump signals at 800 nm can be pumped in counter- propagation along the y-z plane. The non zero second order polarisation of the GaN allows, due again to the $d_{33}$ non linear tensor element , the generation in the x-z plane of a second harmonic field. In this case, such a frequency conversion occurs in a thin layer, thus phase matching conditions are no more strictly mandatory.

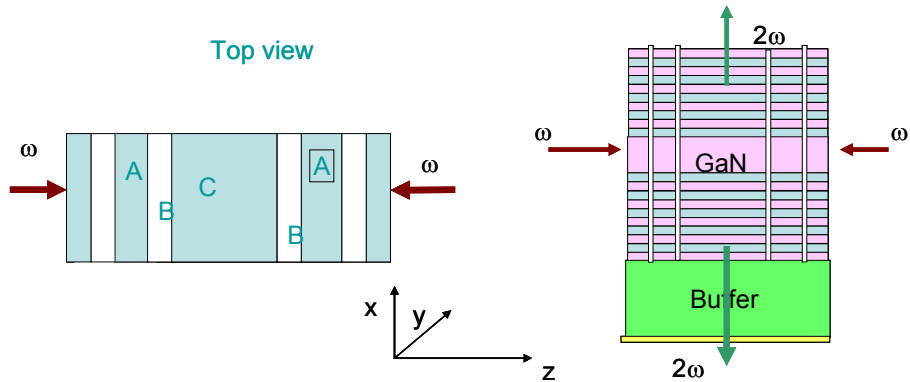

Figure 3. Scheme of the optimised structure in the configuration of SHG experiment.

## 3. FABRICATION OF GAN/ALGAN 1D PHOTONIC CRYSTALS

The fabrication procedure for our GaN/AlGaN 1D-PhC gratings is as follows. First, we grew epitaxial GaN/AlGaN heterostructures by means of metal organic chemical vapour deposition (MOCVD) on sapphire substrates, with to objective of obtaining abrupt and well defined interfaces, together with a good control in the layer thickness.. Figure 1-b shows a cross-section image acquired by scanning electron microscope (SEM) of the grown sample. Both top and bottom mirrors consist of λ/4 layers of epitaxial GaN and $Al_{0.5}Ga_{0.5}N$ (Δn = 0.4), whereas the central cavity simply embeds 1 λ GaN (160 nm thick).

Subsequently, we focused on the definition of our 1D-PhC gratings into the epitaxial samples.This activity involved a number of design and processing steps, each of which required a degree of optimization and checking. Sub micron patterning with sharp and vertical facets on nitride materials is subject of intense investigation by the scientific community, due to the strong binding energy of GaN and AlGaN alloys and to the electron beam writing on such a low conductivity material. The first step was to optimize the e-beam writing process on the GaN/AlGaN samples. To define the geometry of our 1D-PhC we choose a double layer PMMA resist, which is a positive resist and allows to achieve a high quality *lift-off* process needed for defining metallic mask on GaN/AlGaN samples. In particular, we used a thickness of 500 nm of bi-layer resist (PMMA/MMA). As GaN/AlGaN samples are not conductive, before the deposition of the bi-layer resist, we covered them with 5nm of Cr layer in order to avoid the charge effect which causes the deflection of the e-beam during the exposure leading up to the distortion of the pattern.

The 1D-PhC pattern was written using a Raith150 e-beam lithography system operating at 30 kV. Figure 4 reports a SEM image of the pattern defined after e-beam writing.

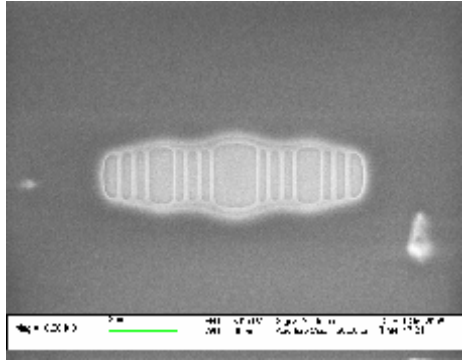

Figure 4. SEM picture of 1D PhC microcavity after e-beam writing.

Following, ~ 170 nm of Cr mask were deposited by thermal evaporator, but after the lift-off, the pattern resulted damaged as shown in figure 5-a. We attributed this damage to a bad adherence of the Cr layer to the GaN/AlGaN surface. Indeed, by replacing the Cr layer with a Titanium one, a better lift off result was achieved as reported in figure 5-b. The final improvement needed to achieve high aspect ratio nanostructures, has been obtained by increasing the thickness of the bi-layer resist up to 700nm. With these conditions we obtained a high quality lift-off as shown in figure 5-c.

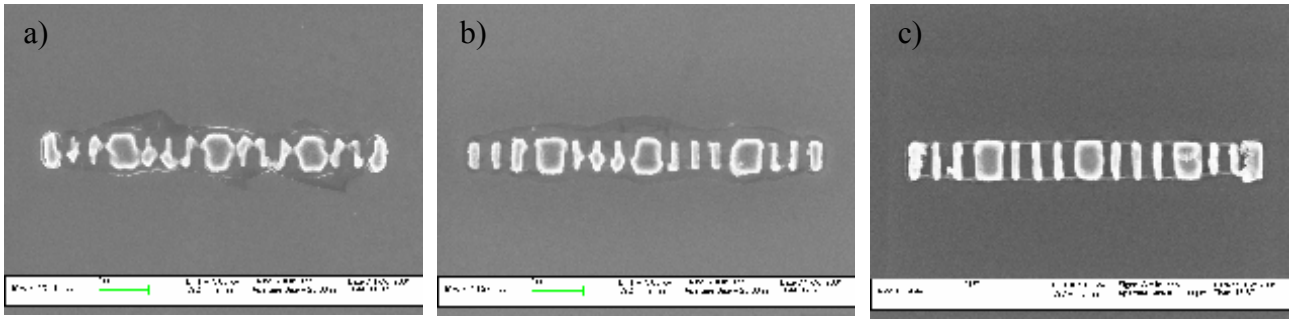

Figure 5. SEM image of 1D-PhC microcavity after lift-off with a) a Cr adherence layer; b) a Ti adherence layer and a thickness of bi-layer resist of 500nm; c) a Ti adherence layer and a thickness of bi-layer resist of 700nm.

At the same time, deep etching processes with different conditions have been performed to transfer these patterns into the GaN/AlGaN heterostructures. By exploiting Inductively Coupled Plasma (ICP) etching, we transferred 1D-PhC patterns from the ~ 170 nm Cr metal etch mask into GaN/AlGaN heterostructures for a depth of about 1μm. The inductively driven source at 13.56 MHz provides a good discharge efficiency because only a smaller part of voltage is dropped across the plasma sheath, which is a thin positively charged layer, and the loss of ion energy is much smaller than capacitive coupling. This efficient plasma discharge makes the plasma density larger than the one in a comparable RIE process. Therefore, the ICP-RIE method is particularly suitable to drill nanometer-scale slots efficiently due to its high etch directionality as we have experimented in this work. Our main task was to achieve slots with vertical and smooth sidewalls which are desirable for fabrication of high quality photonic crystals. By carefully balancing the chemical and physical etching components, depending on coil power, platen power, pressure and gas flow, vertical smooth sidewalls have been obtained. Generally, etching of nitride materials is done using gas mixtures of Ar, $N_2$ and $Cl_2$ [ref. Karouta]. We studied the etch mechanism for our GaN/AlGaN heterostructures in $N_2$:$Ar^+$: $SiCl_4$ plasmas, with special attention for the beneficial role of $SiCl_4$. An example of well balanced etching of nanometer features obtained in GaN/AlGaN material is reported in figure 7. This etching profile has been obtained for a coil power of 600 W, a $SiCl_4$ flow of 20 sccm, a $N_2$ flow of 20 sccm, an $Ar^+$ flow of 5 sccm and a process pressure of 1mTorr. In particular, figure 6-a shows the profile of the GaN/AlGaN 1D-PhC grating, whereas figure 6-b shows the source block with a lateral size of 1 μm.

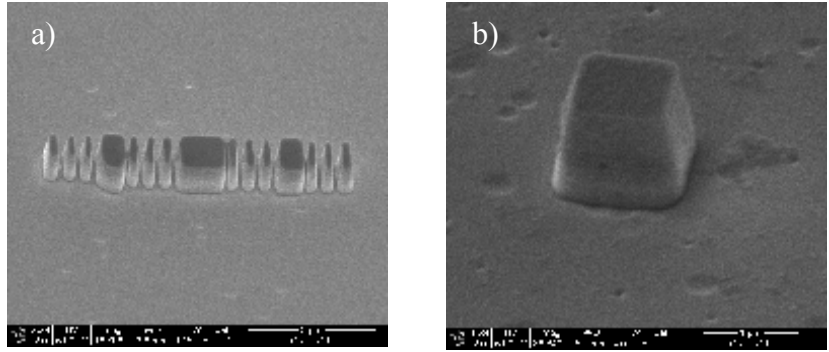

Figure 6. SEM image of the GaN sample etched under the coil power of 600 W, $SiCl_4$ flow of 20 sccm, $N_2$ flow of 20 sccm, $Ar^+$ flow of 5 sccm and process pressure of 1mTorr. a) 1D-PhC grating consisting of slots 200 nm large and 1 µm deep; b) Source block with a lateral size of 1 µm.

A further increase of $Ar^+$ and $N_2$ flow (257 sccm and 25 sccm, respectively) lead to a minimisation the observed pyramid profile (Figure 6) and to nearly vertical sidewalls for nanostructure (200nm wide) etched down to 1 µm with a mask/semiconductor selectivity of 1:7 (Figure 7).

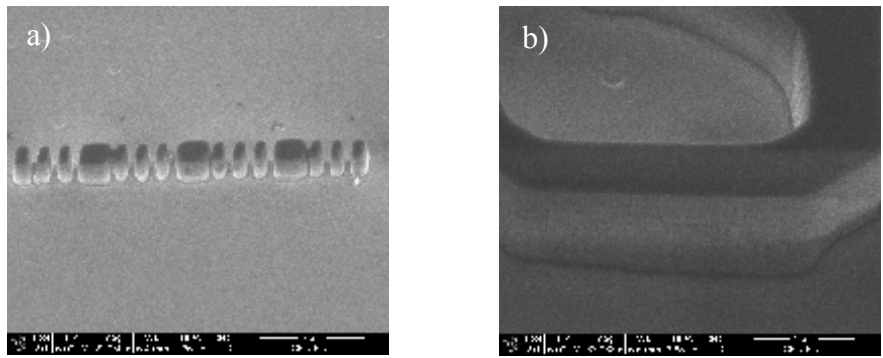

Figure 7. SEM image of the GaN sample etched under the coil power of 600 W, $SiCl_4$ flow of 20 sccm, $N_2$ flow of 25 sccm, $Ar^+$ flow of 7 sccm and process pressure of 1mTorr. a) 1D PhC grating consisting of slots 200 nm large and 0.7 µm deep . b) Shape of the obtained vertical profile.

All the SEM images shown above are related to the first designed 1D PhC structure, as reported in Figure 1. The planar configuration according to the model depicted in fig.2 and fig 3, for side pumping and consisting of a grating with period of 230 nm and slots width of 100 nm is shown in Figure 8.

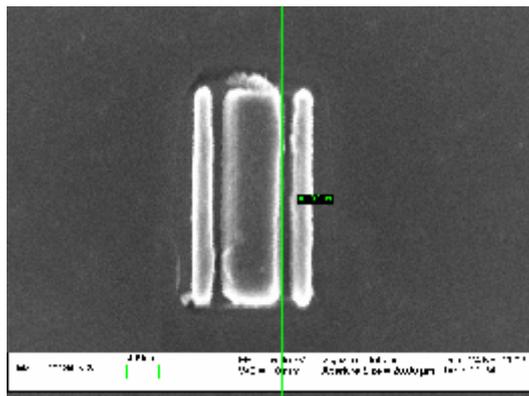

Figure 8. SEM picture of second 1D PhC microcavity after e-beam writing and lift-off processes.

## 4. CONCLUSIONS

We developed a reliable technology to fabricate GaN/AlGaN 1D-PhC microcavities with high aspect ratio (15:1) for the propagation of coherent light generated by non linear optical processes. We used a sample with a GaN layer embedded between two DBRs consisting of AlGaN/GaN multilayers, on sapphire substrate, designed to generate a $\lambda$= 800 nm frequency down-converted signal from an incident pump signal at $\lambda$= 400 nm. The heterostructure was epitaxially grown by MOCVD technique and integrates a properly designed 1D-PhC grating, which amplifies the signal by exploiting the double effect of cavity resonance and non linear light enhancement. 1D-PhC grating with well controlled sizes (periods of 150nm, 230nm and 400nm) were achieved by combing a high resolution e-beam writing with a deep ICP etching in $SiCl_4$. The high accuracy shown from the developed technology is a mandatory feature for the propagation of coherent light generated by non linear optical processes.